\documentclass[10pt,conference]{IEEEtran}
\IEEEoverridecommandlockouts

\def\BibTeX{{\rm B\kern-.05em{\sc i\kern-.025em b}\kern-.08emT\kern-.1667em\lower.7ex\hbox{E}\kern-.125emX}}
   
\newif\ifPATENT
\newif\ifICPP

\ICPPtrue
\PATENTfalse

\usepackage{amsmath}
\usepackage{amsfonts}
\usepackage{amssymb}

\usepackage{paralist}
\usepackage{float}
\usepackage{tikz}
\usepackage{comment}
\usepackage{graphicx,pifont,color}
\usepackage{hhline}
\usepackage{url}
\usepackage[hidelinks]{hyperref}
\usepackage{multirow}

\usepackage{pgfplots}
\pgfplotsset{compat=1.10}
\usetikzlibrary{positioning,calc}
\usetikzlibrary{arrows.meta,chains,decorations.pathreplacing,scopes,positioning}
\usetikzlibrary{calc,shapes.multipart,chains,arrows}
\usepackage{lhelp}
\usepackage{listings}
\usepackage{xcolor,colortbl}
\usepackage{tcolorbox}

\usepackage{mathtools}
\usepackage{bm}
\usepackage{esvect}

\newtheorem{definition}{Definition}

\newtheorem{example}{Example}

\newtheorem{observation}{Observation}
\newtheorem{remark}{Remark}

\def\Q{\mbox{${\mathbb Q}$}}

\newcommand{\cupti}{\mbox{\sc CUPTI}}
\def\cuda{\mbox{\sc CUDA}}

\def\mwpcwp{\mbox{MWP-CWP}}

\newcommand{\nvprof}{\mbox{\sc nvprof}}
\newcommand{\dimThree}{\mbox{\texttt{dim3}}}

\newcommand{\hidetext}[1]{\mbox{ \ }}

\newcommand{\fixed}[2]{#1}

\definecolor{wp}{rgb}{0.129,0.478, 0.678}
\definecolor{wpbright}{rgb}{0.26,0.55, 0.79}
\definecolor{yellowBright}{HTML}{FFD247}
\definecolor{orangeBright}{HTML}{FF7E47}
\definecolor{gray1}{HTML}{CEDBE1}
\definecolor{gray2}{HTML}{DDE6EA}
\definecolor{gray3}{HTML}{E3E6EB}
\definecolor{darkblue1}{HTML}{2375B9}
\definecolor{darkblue2}{HTML}{3D577A}
\definecolor{lightblue1}{HTML}{65A4D9}
\definecolor{lightblue2}{HTML}{94C3E9}
\definecolor{red1}{HTML}{E4001B}
\definecolor{red2}{HTML}{8F2831}
\definecolor{lime1}{HTML}{AEBD63}
\definecolor{navy}{HTML}{000063}
\definecolor{purple1}{HTML}{8F2831}
\definecolor{contrast}{HTML}{CE0000}
\definecolor{positive}{HTML}{0D900F}
\definecolor{negative}{HTML}{FF0000}

\newcommand{\removed}[1]{}

\newif\ifready
\readyfalse

\newif\ifstable
\stabletrue

\newif\ifnewversion
\newversiontrue

\makeatletter
\DeclareRobustCommand*\cal{\@fontswitch\relax\mathcal}
\makeatother

\newif\ifready
\readyfalse

\newif\ifshortversion
\shortversiontrue

\newif\ifstable
\stabletrue

\newif\ifnewversion
\newversiontrue

\title{
KLARAPTOR: A Tool for Dynamically Finding Optimal Kernel Launch Parameters Targeting CUDA Programs\\
\thanks{This work has been supported in part by IBM Canada Ltd (CAS project 880) and in part by
NSERC of Canada (CRD grant CRDPJ500717-16, award PGSD3-535362-2019).}
}

\author{
\IEEEauthorblockN{Alexander Brandt\IEEEauthorrefmark{1},
Davood Mohajerani\IEEEauthorrefmark{1},
Marc Moreno~Maza\IEEEauthorrefmark{1},
Jeeva Paudel\IEEEauthorrefmark{2} and
Linxiao Wang\IEEEauthorrefmark{1}}
\IEEEauthorblockA{\IEEEauthorrefmark{1}Department of Computer Science, University of Western Ontario, London, Canada}
\IEEEauthorblockA{\IEEEauthorrefmark{2}IBM Canada Software Labratory, Markham, Canada}
Email: abrandt5@uwo.ca, dmohajer@uwo.ca, moreno@csd.uwo.ca, pjeeva01@ca.ibm.com, lwang739@uwo.ca
}

\begin{document}

\maketitle

\begin{abstract}
\ifICPP
In this paper we present KLARAPTOR (Kernel LAunch parameters RAtional Program estimaTOR),
a new tool built on top of the LLVM Pass Framework and NVIDIA {\cupti} API
to dynamically determine the optimal values of kernel launch parameters of a {\cuda} program ${\cal P}$.
To be precise, we describe a novel technique to statically build (at the compile time of ${\cal P}$) 
a so-called rational program ${\cal R}$. Using a performance
prediction model, and knowing particular data and hardware parameters
of ${\cal P}$ at runtime, the program ${\cal R}$
can automatically and dynamically determine the values 
of launch parameters of ${\cal P}$ that will yield optimal performance.
\fi 
Our technique can be applied to parallel programs in general,
as well as to generic performance prediction models 
which account for program and hardware parameters. 
We are particularly interested in programs targeting manycore accelerators.
We have implemented and successfully tested our technique in the context of
GPU kernels written in {\cuda} using the {\mwpcwp} performance prediction model.

\end{abstract}


\section{Introduction}
\label{sec:intro}

Programming for high-performance parallel computing is a notoriously
difficult task. Programmers must be conscious of many factors impacting
performance including scheduling, synchronization, and data locality.
Of course, program code itself impacts the program's performance, however, 
there are still further \textit{parameters} which are independent from the code
and greatly influence performance. 
For parallel programs \removed{on multiprocessors} three types of parameters influence performance:
\begin{inparaenum}[(i)]
	\item {\em data parameters}, such as input data and its size,  
	\item {\em hardware parameters}, such as cache capacity and number of 
	available registers, and
	\item {\em program parameters}, such as granularity of tasks and 
	the quantities that characterize how tasks are
	mapped to processors (e.g. dimension sizes of a thread block for a {\cuda} kernel).
\end{inparaenum}

Data and hardware parameters are independent from program parameters
and are determined by the needs of the user and available hardware
resources.  Program parameters, however, are intimately related to
data and hardware parameters. The choice of program parameters can
largely influence the performance of a parallel program, even up to an
order of magnitude difference (see Section~\ref{sec:performance}).
Therefore, it is critical to determine optimal values of program
parameters that yield the best program performance for a given set of
hardware and data parameter values.

In the CUDA programming model the kernel launch parameters,
and thus the size and shape of thread blocks, greatly impact performance.
This should be obvious considering that the memory accesses pattern of threads
in a thread block can depend on the block's dimension sizes. 
The same could be said about multithreaded programs on CPU where
parallel performance depends on task granularity and number of threads.
We dedicate, however, this paper to the discussion of 
GPU programs where optimizing performance is much more difficult.
An important consequence of the impact of kernel launch parameters on performance
is that optimal thread block format (that is, dimension sizes)
for one GPU (Graphics Processing Unit) architecture may 
not be optimal for another, as illustrated
in~\cite{DBLP:journals/tjs/TorresGL13}.
This emphasizes not only the impact of hardware parameters on program parameters, 
but also the need for performance portability.
That is to say, enabling users to efficiently execute the same parallel program
on different architectures that belong to the same hardware platform.

\begin{figure}[hbt]
	\includegraphics[width=\columnwidth,trim={2em, 2em, 2em, 2em}, clip]{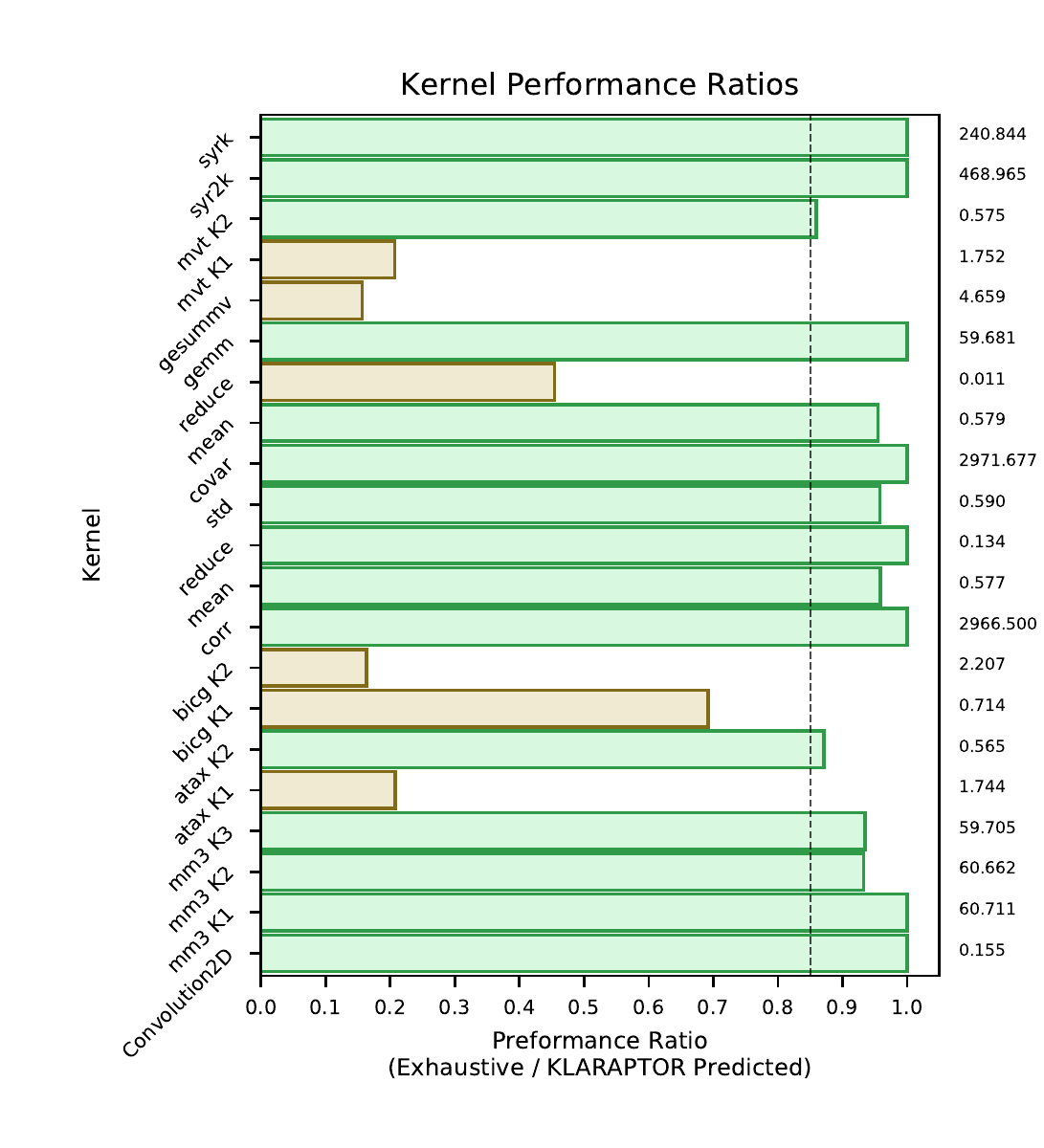}
	\caption{Comparing kernel execution times with optimal launch parameters as determined by an exhaustive search versus KLARAPTOR's prediction.
	Kernels are part of the \texttt{PolyBench/GPU} benchmark suite and executed on a GTX 1080Ti with
	a data size of $N=2048$. Ratios greater than 85\% are considered good.
	Absolute timings given next to each bar are in ms.
	Due to memory limitations not all benchmarks could not run for $N=2048$.}
        \label{fig:kernelperf2048}
\end{figure}

In this paper, we describe the development of KLARAPTOR (Kernel LAunch
parameters RAtional Program estimaTOR), a tool
for automatically and
dynamically determining the values of  {\cuda} kernel launch parameters which
optimize the kernel's performance, for each kernel launch
independently.  That is to say, based on the actual data and target
device of a kernel invocation.
The source code of KLARAPTOR is freely available at
\textcolor{navy}{\href{https://github.com/orcca-uwo/KLARAPTOR}{https://github.com/orcca-uwo/KLARAPTOR}}.

The accuracy of KLARAPTOR's optimal prediction is illustrated in
Fig.~\ref{fig:kernelperf2048} where kernel execution times 
are compared between KLARAPTOR's choice of thread block configuration
and the optimal thread block configuration determined by an exhaustive search.

KLARAPTOR applies to {\cuda} a generic and novel technique, also described herein in Section~\ref{sect:steps}, 
to statically build a so-called \textit{rational program} 
which is then used dynamically to determine optimal program parameters
for a given multithreaded program on specific data and hardware parameters.
The key principle underpinning the proposed technique can be
summarized as follows.  

In most performance predictions models, \textit{high-level performance metrics}, such
as execution time, memory consumption, and hardware occupancy, are
piece-wise rational functions (PRFs) of \textit{low-level performance metrics}, such as 
memory bandwidth, cache miss rate, and the number of clock cycles per
instruction.
These lower-level metrics are themselves PRFs of
program, data, and hardware parameters. As such,
a high-level performance metric can be estimated by a piece-wise
rational function of the program, data, and hardware parameters. 
For a fixed machine, the hardware parameters are also fixed
and performance metrics depend only on program and data parameters.  
Henceforth, we regard a computer program that computes such a piece-wise rational
function as a {\em rational program},
a technical notion fully defined in Section~\ref{sec:foundations}.

In most cases, the values of the data parameters are only given at
runtime, which makes it difficult to determine optimal values of the
program parameters at an earlier stage. On another hand, a bad choice
of program parameters can have drastic consequences.  Hence, it is
crucial to be able to determine the optimal program parameters at
runtime without much overhead added to the program execution. This is
precisely the intention of the approach proposed in this paper.


\subsection{Contributions}

The goal of this work is to determine values of program parameters which
optimize a multithreaded program's performance.  Towards that goal,
the method by which such values are found must be
receptive to changing data and changing  hardware parameters. Our contributions
encapsulate this requirement through the dynamic use of a rational
program.  Our specific contributions include:
\begin{enumerate}[(i)]
	\item a technique for devising a mathematical expression in the form of a {\em rational program} to evaluate a high-level performance metric from a set of program and data parameters;
	\item KLARAPTOR, a tool implementing the rational program technique to dynamically optimize {\cuda} kernels by choosing optimal launch parameters; and
	\item an empirical and comprehensive evaluation of our tool on kernels from the \texttt{Polybench/GPU} benchmark suite.
\end{enumerate}

\subsection{Related Works}
\label{sec:relatedworks}
The {\em Parallel Random Access Machine}
(PRAM) model~\cite{DBLP:journals/siamcomp/StockmeyerV84,Gibbons:1989:MPP:72935.72953},
including PRAM models tailored to GPU code analysis
such as 
TMM~\cite{DBLP:journals/fgcs/MaAC14} and  MCM~\cite{DBLP:conf/parco/HaqueMX15}
analyze the performance of parallel programs at an abstract level.
More detailed GPU performance models are proposed such as MWP-CWP 
by Hong et al \cite{DBLP:conf/isca/HongK09, DBLP:conf/ppopp/SimDKV12}, which estimates the execution time of GPU kernels based on the profiling information of the kernels.

In the context of improving {\cuda} program performance, other research groups have
used techniques such as loop transformation \cite{DBLP:conf/ics/BaskaranBKRRS08},
auto-tuning~\cite{grauer2012auto,Khan:2013:SAC:2400682.2400690,DBLP:books/daglib/p/SatoTKK10,DBLP:conf/icpp/KurzakTGAD19},
dynamic instrumentation~\cite{kistler2003continuous}, or
a combination of the latter two~\cite{song2015automated}.
Auto-tuning techniques have achieved great results
in projects such as 
ATLAS~\cite{DBLP:conf/ppsc/WhaleyD99},
FFTW~\cite{DBLP:conf/icassp/FrigoJ98}, and
SPIRAL~\cite{DBLP:journals/ijhpca/PuschelMSXJPVJ04}
in which multiple kernel versions are generated 
\textit{off-line} and then applied and refined  \textit{on-line}
once the runtime parameters are known.

Although much research has been devoted to optimizing a kernel
algorithmically, previous works such
as~\cite{DBLP:conf/cascon/ChenCKMX15} and~\cite{DBLP:journals/tjs/TorresGL13} suggest that kernel launch
parameters (i.e. thread block configurations)
have a large impact on performance and must be considered
as a target for optimization.
In~\cite{DBLP:conf/icpp/LimNM17}
the authors claim to have a static analysis method which does not
require running the code for determining the best configuration for
{\cuda} kernels. 
Through static code analysis, IPC (instructions per clock-cycle)
is estimated, however, 
there is no analysis of memory access patterns.
The authors assume 
``the execution time of a {\cuda} program is proportional to the
input problem size $N$'' \cite[pp.527]{DBLP:conf/icpp/LimNM17}.
This is, obviously, a very strong assumption, and impractical
for real world applications where even simple operations such as
matrix multiplication are not proportional to their input size.

In \cite{liu2009cross}, the authors present an input-adaptive GPU code
optimization framework G-ADAPT, which uses statistical learning to
find a relation between the input sizes and the thread block sizes.
At runtime or linking time, the framework predicts the best block size
for a given input size using the linear model obtained from compile
time.  This approach only considers the total size of the thread
blocks and not their configuration.  Meanwhile, the authors
of~\cite{DBLP:books/daglib/p/SatoTKK10} use a linear regression model
to predict optimal thread block configurations (that is, dimension
sizes and not just the total size).  However, like the authors
of \cite{DBLP:conf/icpp/LimNM17}, they assume that the execution time
of a kernel is proportional to its data size. Moreover, only two
benchmarks are provided.
In~\cite{DBLP:conf/icpp/GarveyA15},
machine learning techniques are used in combination with auto-tuning
to search for optimal configurations of OpenCL kernels.
Their examples are limited to stencil computations.

\subsection{Structure of the Paper}

The remainder of this paper is organized as follows.
Section~\ref{sec:foundations} formalizes and exemplifies the notion of rational programs
and their relation to piece-wise rational functions and performance prediction.
Section~\ref{sec:mainresults} highlights our two main results.
Firstly, that rational programs can be determined as a mathematical expression for
evaluating a performance metric from program and data parameters and, secondly, our success in applying this technique
to {\cuda} kernels via the KLARAPTOR tool. 

The remaining sections 
examine our generic technique for building and using a rational program 
to estimate program performance (Section~\ref{sect:steps}),
the specifics of the implementation of that technique as KLARAPTOR (Section~\ref{sec:embodiment}),
and further experimentation and evaluation of our implementation (Section~\ref{sec:performance}).
We draw conclusions and explore future work in Section~\ref{sec:conclusion}.


\section{Theoretical Foundations}
\label{sec:foundations}

Let ${\cal P}$ be a multithreaded program to be executed on a targeted
multiprocessor.  By fixing the target architecture, the hardware
parameters, say, ${\bm H} = \left(H_1, \ldots, H_h\right)$ then become
fixed and we can assume that the performance of ${\cal P}$ depends
only on data parameters ${\bm D} = \left(D_1, \ldots, D_d\right)$ and
program parameters ${\bm P} = \left(P_1, \ldots, P_p\right)$. Also,
an optimal choice of ${\bm P}$ depends on a specific choice of ${\bm D}$.
For example, in programs targeting GPUs 
the parameters ${\bm D}$ are typically
dimension sizes of data structures, like arrays, 
while ${\bm P}$ typically specify the formats
of thread blocks.

Let ${\cal E}$ be a high-level performance metric for ${\cal P}$ that we want to
optimize.  More precisely, given the values of the data parameters
${\bm D}$, the goal is to find values of the program parameters
${\bm P}$ such that the execution of ${\cal P}$ optimizes
${\cal E}$.  
Performance prediction models attempt to estimate ${\cal E}$ from a
combination of ${\bm P}$, ${\bm D}$, ${\bm H}$, and some model- or
platform-specific low-level metrics ${\bm L} = (L_1, \ldots,
L_\ell)$.  It is natural to assume that these low-level performance
metrics are themselves combinations of ${\bm P}$, ${\bm D}$, ${\bm
  H}$.  This is an obvious observation from models based on PRAM such
as TMM~\cite{DBLP:journals/fgcs/MaAC14} and
MCM~\cite{DBLP:conf/parco/HaqueMX15}.
Hence, regardless of the particular performance prediction model, we may 
drop direct dependency on ${\bm L}$ and 
look to estimate ${\cal E}$ exclusively from some (unknown) combination of 
program and data parameters.

To address our goal, we compute a mathematical expression,
parameterized by data and program parameters,
in the format of a {\em rational program} ${\cal R}$ at compile-time. 
At runtime, given the specific values of ${\bm D}$, we can efficiently evaluate ${\cal E}$ 
using ${\cal R}$.
After that, we can feasibly pick a value of ${\bm P}$ that optimizes ${\cal E}$, and feed that to ${\cal P}$ for execution.
We refer to such ${\bm P}$ which optimizes ${\cal E}$
as \textit{optimal values of the program parameters}.
This method is detailed in Section~\ref{sect:steps}.

\subsection{Rational Programs}
\label{sect:ratprog}

Let $X_1, \ldots, X_n, Y$ be pairwise different variables\footnote{Here
variable refers to both the algebraic meaning of a 
polynomial variable and the programming language concept.}.
Let ${\cal S}$ be a sequence  of three-address code (TAC~\cite{Aho:1986:CPT:6448})
instructions such that the set of 
the variables that occur in ${\cal S}$ 
and are never assigned a value
by an instruction of ${\cal S}$ is exactly $\{ X_1, \ldots, X_n \}$.

\begin{definition}
\label{defi:rationalprogram}
We say that the sequence ${\cal S}$ is 
{\em rational} if every
arithmetic operation used in ${\cal S}$ is either an addition, a subtraction,
a multiplication, or a comparison 
of integer numbers
in either fixed or arbitrary precision.
Moreover, we say that the sequence ${\cal S}$ is 
a {\em rational program} in $X_1, \ldots, X_n$ evaluating $Y$
if the following two conditions hold:
\begin{enumerateshort}
\item ${\cal S}$ is rational, and
\item  after specializing each of $X_1, \ldots, X_n$ to an arbitrary 
      integer value in ${\cal S}$,
      the execution of the 
      specialized sequence ${\cal S}$ always terminates and the last
      executed instruction assigns an integer value to $Y$.
\end{enumerateshort}
\end{definition}

\newif\ifshortremarks
\shortremarkstrue

\ifshortremarks
It is worth noting that the above 
definition can easily be extended to include
Euclidean division, the integer part operations floor and ceiling,
and arithmetic over rational numbers.
For Euclidean division one can write a rational program
evaluating the quotient $q$ of integer $a$ by $b$, 
leaving the remainder $r$ to be simply calculated as $a - qb$.
Then, floor and ceiling can be computed via Euclidean division.
Rational numbers and their associated arithmetic is easily implemented using
only integer arithmetic.

Notice then, that by adding these operations to 
Definition~\ref{defi:rationalprogram}, the 
class of rational programs  does not change. Thus, adding rational numbers,
Euclidean division, and integer part operations to
the definition of a rational sequence yields an equivalent definition of a rational program.
We adopt this definition henceforth. 

\subsection{Rational Programs as flowcharts}

For any sequence ${\cal S}$ of computer program 
instructions, it is convenient to associate  ${\cal S}$
with a {\em control flow graph} (CFG).
In the CFG of ${\cal S}$, the nodes are the {\em basic blocks} of ${\cal S}$,
that is, the sub-sequences of ${\cal S}$  such that 
\begin{enumerate}[(i)]
	\item  each instruction except the last one is not a branch, and
	\item which are maximum in length with property (i).
\end{enumerate}
Moreover, in the CFG of ${\cal S}$, 
there is a directed edge from a basic block $B_1$ to a basic block
$B_2$ whenever, during the execution of ${\cal S}$, one can jump from
the last instruction of $B_1$ to the first instruction of $B_2$.
Recall that  a {\em flowchart} is another
graphic representation of a sequence of computer program  instructions.
In fact, CFGs can be seen as particular flowcharts.

If, in a given flowchart ${\cal C}$, every arithmetic
operation occurring in every (process or decision) node is
either an addition, subtraction, multiplication, or comparison of integers
in either fixed or arbitrary precision
then ${\cal C}$ is the flowchart of a rational sequence of computer
program instructions.
Therefore, it is meaningful to depict rational programs
using flowcharts. 
Two meaningful examples of rational programs 
are presented in the following section while 
Fig.~\ref{fig:occupancysimpleflowchart} 
depicts one of them as a flowchart.

\else
The above definition calls for a few natural remarks.

\begin{remark}
{\em
\label{remark:div}
One can easily extend Definition~\ref{defi:rationalprogram}
by allowing the use of the Euclidean division for integers, 
in both fixed and arbitrary precision.
Indeed, one can write a rational program evaluating the 
integer quotient $Q$ of a signed integer $A$ by a signed integer $B$.
Then, the remainder $R$ in the Euclidean division
of $A$ by $B$ is simply $A - Q B$.
}
\end{remark}

\begin{remark}
{\em
One can also extend Definition~\ref{defi:rationalprogram}
by allowing addition, subtraction,
multiplication and  comparison ($=$, $<$) of rational numbers
in arbitrary precision.
Indeed, each of these operations can easily be implemented
by rational programs using addition, subtraction,
multiplication and  comparison ($=$, $<$) for integer numbers
in arbitrary precision.
}
\end{remark}

\begin{remark}
\label{remark:floor}
{\em
Next, one can extend Definition~\ref{defi:rationalprogram}
by allowing the {\em integer part} operations
$F \longmapsto \lceil F \rceil$
and $F \longmapsto \lfloor F \rfloor$
where $F$ is an arbitrary rational number.
Indeed, for a rational number $A/B$ 
(where $A$ and $B$ are signed integers)
the integer parts $\lceil A/B \rceil$
and $\lfloor A/B \rfloor$ can be computed
by performing the Euclidean division
of $A$ by $B$.
Consequently, one can allow TAC instructions
of the forms $Q = \lceil A/B \rceil$
and $Q = \lfloor A/B \rfloor$, where $=$ 
denotes the assignment as in the C programming
language.
}
\end{remark}

\begin{remark}
{\em
Extending Definition~\ref{defi:rationalprogram} according to
Remarks \ref{remark:div}--\ref{remark:floor} does not change
the class of rational programs. Thus, adding Euclidean division, 
rational number arithmetic and integer part computation to the 
definition of a rational sequence yields an equivalent definition
of rational program. We adopt this definition henceforth.
}
\end{remark}

\begin{remark}
{\em
Recall that it is convenient 
to associate any sequence ${\cal S}$ of computer program 
instructions
with a {\em control flow graph} (CFG).
In the CFG of ${\cal S}$, the nodes are the {\em basic blocks} of ${\cal S}$,
that is, the sub-sequences of ${\cal S}$  such that 
\begin{inparaenum}[(i)]
  \item  each instruction except the last one is not a branch, and
   \item which are maximum in length with property $(i)$.
 \end{inparaenum}
Moreover, in the CFG of ${\cal S}$, 
there is a directed edge from a basic block $B_1$ to a basic block
$B_2$ whenever, during the execution of ${\cal S}$, one can jump from
the last instruction of $B_1$ to the first instruction of $B_2$.
Recall also that  a {\em flowchart} is another
graphic representation of a sequence of computer program  instructions.
In fact, CFGs can be seen as particular flowcharts.

If, in a given flowchart ${\cal C}$, every arithmetic
operation occurring in every (process or decision) node is
either an addition, subtraction, multiplication, or comparison, for integers
in either fixed or arbitrary precision (or any other operation
as explained in Remarks~\ref{remark:div}-\ref{remark:floor}),
then ${\cal C}$ is the flowchart of a rational sequence of computer
program instructions.
Therefore, it is meaningful to depict rational programs
using flowcharts. An example is given by 
Fig.~\ref{fig:occupancysimpleflowchart}.
}
\end{remark}
\fi

\tikzset{%
	>={Latex[width=2mm,length=2mm]},
	base/.style = {rectangle, rounded corners, draw=black,
		minimum width=4cm, minimum height=1cm,
		text centered, font=\sffamily},
	activityStarts/.style = {base, fill=blue!30},
	startstop/.style = {base, fill=red!30},
	activityRuns/.style = {base, fill=green!30},
	process/.style = {base, minimum width=2.5cm, fill=blue!15,
		font=\ttfamily},
	ifstatement/.style = {base, diamond, aspect=2.5, scale=0.7, font=\Large},
}
\usetikzlibrary{shapes.geometric}

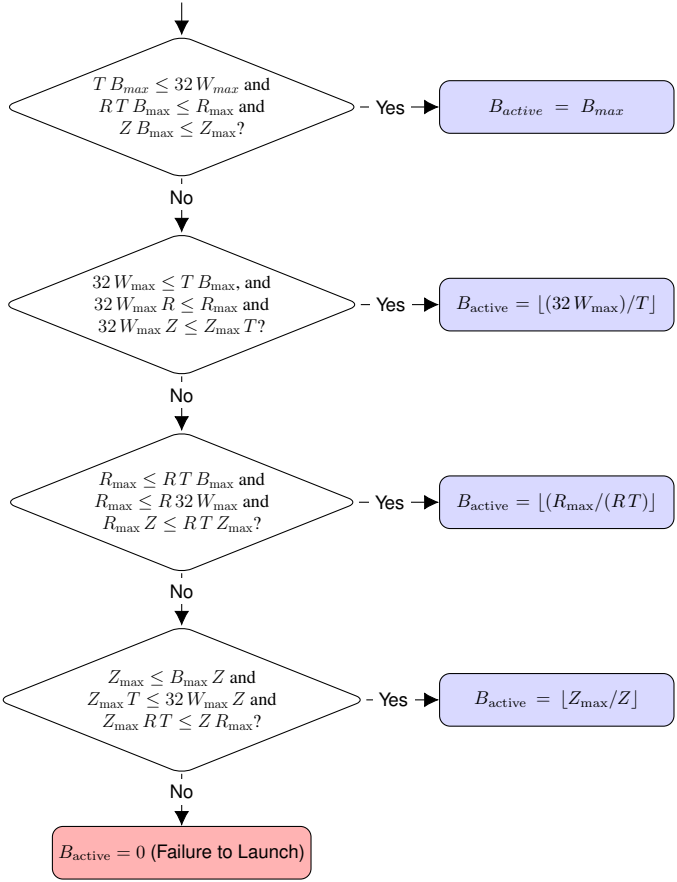
\begin{figure}
	\centering
\begin{tikzpicture}[node distance=1.5cm,scale=.5,
    every node/.style={fill=white,scale=.7, font=\sffamily}, align=center]
  
  \node (activemax) [ifstatement] {$T \, B_{max} \leq 32 \, W_{max}$ and\\ $R \,  T \,  B_{\rm max} \leq R_{\rm max}$ and\\ $Z \,  B_{\rm max} \leq Z_{\rm max}$?};
  
  \node (activewmax) [ifstatement, below of=activemax, yshift=-11em] {$32 \,  W_{\rm max} \leq T \,  B_{\rm max}$, and\\ $32 \,  W_{\rm max} \,  R \leq R_{\rm max}$ and\\ $32 \,  W_{\rm max} \,  Z \leq Z_{\rm max} \,  T$?};
  
  \node (activermax) [ifstatement, below of=activewmax, yshift=-11em] {$R_{\rm max} \leq R \,  T \,  B_{\rm max}$ and\\ $R_{\rm max} \leq R \,  32 \,  W_{\rm max}$ and\\ $R_{\rm max} \,  Z \leq R \,  T \,  Z_{\rm max}$?};

  \node (activezmax) [ifstatement, below of=activermax, yshift=-11em] {$Z_{\rm max} \leq B_{\rm max} \,  Z$ and\\ $Z_{\rm max} \,  T \leq 32 \,  W_{\rm max} \,  Z$ and\\ $Z_{\rm max} \,  R \,  T \leq Z \,  R_{\rm max}$?};

  \node (bmax) [process, right of=activemax, xshift=16em, text width=12em] {$B_{active} = B_{max}$};
  \node (wmax) [process, right of=activewmax, xshift=16em, text width=12em] {$B_{\rm active} =  \lfloor( 32 \,  W_{\rm max})/T  \rfloor$};
  \node (rmax) [process, right of=activermax, xshift=16em, text width=12em] {$B_{\rm active} =  \lfloor( R_{\rm max}/(R \,  T) \rfloor$};
  \node (zmax) [process, right of=activezmax, xshift=16em, text width=12em] {$B_{\rm active} =  \lfloor Z_{\rm max}/Z \rfloor$};
      
  \node (fail) [startstop, below of=activezmax, yshift=-4em] {$B_{\rm active} = 0$
(Failure to Launch)};

  \draw[->] (activemax) -- node[] [yshift=0.45em]{No} (activewmax);
  \draw[->] (activewmax) -- node[] [yshift=0.45em]{No} (activermax);
  \draw[->] (activermax) -- node[] [yshift=0.45em]{No} (activezmax);
  \draw[->] (activezmax) -- node[] [yshift=0.45em]{No} (fail);  
  
  \draw[->] (activemax) -- node[] [xshift=-0.5em]{Yes} (bmax);
  \draw[->] (activewmax) -- node[] [xshift=-0.5em]{Yes} (wmax);
  \draw[->] (activermax) -- node[] [xshift=-0.5em]{Yes} (rmax);
  \draw[->] (activezmax) -- node[] [xshift=-0.5em]{Yes} (zmax);
  
  \draw[->]  ([shift=({0,2.5em})]activemax.north) to (activemax.north);
\end{tikzpicture}
\caption{Rational program (presented as a flow chart) for 
the calculation of hardware occupancy in {\cuda}.}
\label{fig:occupancysimpleflowchart}
\end{figure}

\subsection{Examples}

\begin{example}
\label{ex:cuda}
{\em Hardware occupancy}, as defined 
in the {\cuda} programming model, is a measure of a program's effectiveness in using
the Streaming Multiprocessors (SMs) of a GPU.
Hardware occupancy is calculated from a number of hardware parameters, namely:
\begin{itemizeshort}
\item[-] the maximum number $R_{\rm max}$ of registers per thread block,
\item[-] the maximum number $Z_{\rm max}$ of shared memory words per thread block,
\item[-] the maximum number $T_{\rm max}$ of threads per thread block,
\item[-] the maximum number $B_{\rm max}$ of thread blocks per SM and 
\item[-] the maximum number $W_{\rm max}$ of warps per SM,
\end{itemizeshort}
as well as low-level kernel-dependent performance metrics, namely:
\begin{itemizeshort}
\item[-] the number $R$ of registers used per thread and 
\item[-] the number $Z$ of shared memory words used per thread block,
\end{itemizeshort}
and a program parameter, namely the 
number $T$ of threads per thread block.
The occupancy of a {\cuda} kernel is defined as
the ratio between the number of active warps
per SM and the maximum number of warps per SM, namely
$W_{\rm active} / W_{\rm max}$, where
\begin{equation}
\label{eq:occupancy}
    W_{\rm active} =  \min \left( \lfloor B_{\rm active} T / 32 \rfloor, W_{\rm max}  \right)
\end{equation}
and $B_{\rm active}$ is given as a flowchart
by Fig.~\ref{fig:occupancysimpleflowchart}.
This flowchart shows
how one can derive a rational program computing
$B_{\rm active}$ from $R_{\rm max}$, $Z_{\rm max}$,
$T_{\rm max}$, $B_{\rm max}$, $W_{\rm max}$, $R$, $Z$, $T$.
It follows from Equation (\ref{eq:occupancy})
that $W_{\rm active}$ can also be computed 
by a rational program from $R_{\rm max}$, $Z_{\rm max}$,
$T_{\rm max}$, $B_{\rm max}$, $W_{\rm max}$, $R$, $Z$, $T$.
Finally, the same is true for the occupancy of a {\cuda} kernel
using $W_{active}$ and $W_{max}$.
\end{example}

\begin{example}
\label{ex:mwpcwp}
The {\em execution time} of a GPU kernel is 
a high-level performance metric defined 
in the MWP- model \cite{DBLP:conf/isca/HongK09,DBLP:conf/ppopp/SimDKV12}
which is calculated from hardware parameters including:
\begin{itemizeshort}
\item[-] clock frequency of a SM,
\item[-] the number of SMs on the device, etc.,
\end{itemizeshort}
as well as low-level kernel-dependent performance metrics, including:
\begin{itemizeshort}
\item[-] the total number $\#{\rm Mem\_insts}$
of memory instructions per thread,
\item[-] the total number $\#{\rm Comp\_inst }$ of computation 
instructions per thread, etc.,
\end{itemizeshort}
and a program parameter, namely the 
number $T$ of threads per thread block.
Thus, this high-level metric can be computed as rational program 
of program, data, and hardware parameters. 
Indeed, we have realized such a rational program making us of 
the {\mwpcwp} model and its metrics
in our implementation of KLARAPTOR (See Sections~\ref{sec:mainresult_klaraptor} and \ref{sec:embodiment}).
\end{example}

\subsection{Piece-Wise Rational Functions in Rational Programs}

\begin{observation}
\label{obs:rationality}
\fixed{
Let ${\cal S}$ be a rational program in $X_1, \ldots, X_n$ evaluating $Y$.
Let $s$ be any instruction of ${\cal S}$ other than a branch or an
integer part instruction.  
Hence, this instruction can be of the form
$C = -A$, $C = A + B$, $C = A - B$, $C = A \times B$, where $A$ and
$B$ can be any machine-representable rational numbers.
Let $V_1, \ldots, V_v$ be the variables that are defined
at the entry point of the basic block of  the instruction $s$.
An elementary proof by induction yields the following fact.
There exists a rational function\footnote{Here, rational function is in the
sense of algebra, see~\cite{eisenbud2013commutative} and section \ref{sec:paramest}.} in $V_1,
\ldots, V_v$ that we denote by $f_s(V_1, \ldots, V_v)$
such that $C = f_s(V_1, \ldots, V_v)$ for all possible values of
$V_1, \ldots, V_v$.}
{Should this be written as w.r.t a basic code block rather than a single instruction? No, because the goal is to state this property for those instructions evaluating $Y$.}
From there, one derives the following observation.
There exists a partition ${\cal T} = \{ T_1, T_2, \ldots    \}$ 
of ${\Q}^n$ (where ${\Q}$ denotes the field of rational numbers)
and rational functions $f_1(X_1, \ldots, X_n)$, 
$f_2(X_1, \ldots, X_n)$, \ldots 
such that, if $X_1, \ldots, X_n$ receive respectively 
the values $x_1, \ldots, x_n$, then
the value of $Y$ returned by ${\cal S}$ is one of
$f_i(x_1, \ldots, x_n)$ where $i$ is such that
$(x_1, \ldots, x_n) \in T_i$ holds.
In other words, ${\cal S}$ computes $Y$ as a
{\em piece-wise rational function} \, (PRF).

Example~\ref{ex:cuda} 
shows that the hardware occupancy of a {\cuda}
kernel is given as a piece-wise rational function
in the variables $R_{\rm max}$, $Z_{\rm max}$,
$T_{\rm max}$, $B_{\rm max}$, $W_{\rm max}$, $R$, $Z$, $T$.
Hence, in this example, we have $n = 8$.
Moreover, Fig.~\ref{fig:occupancysimpleflowchart} 
shows that its partition of ${\Q}^n$ contains 5 parts
as there are 5 terminating nodes in the flowchart.
\end{observation}

\section{Main Results}
\label{sec:mainresults}

\subsection{Rational Programs for Performance Prediction}
\label{sec:mainresult_rp}

In Section~\ref{sec:foundations}, we saw how a rational program
could be seen as a piece-wise rational function.
We now show how a rational program can encode, and thus evaluate, a performance
prediction model using only program and data parameters. 
A technique for constructing such a rational program is given in Section~\ref{sect:steps}.

\begin{observation}
\label{obs:faisability}
We assume that the high-level performance metric ${\cal E}$
is computable as a rational program depending on program, data, and 
hardware parameters,
and some low-level performance metrics.
Then, it is natural to assume that these low-level metrics
also depend only rationally (as opposed to some more complex dependency)
on program, data, and hardware parameters. 
Hence, we replace the direct dependency of ${\cal E}$ on low-level metrics
with a further dependency on the program, data, and hardware parameters.
When applied to a multithreaded program 
to be executed on a particular multiprocessor, the hardware parameters
become specialized and we can assume that ${\cal E}$ can be expressed
as a rational program depending only on program and data parameters. 
\end{observation}

Observations~\ref{obs:rationality} and \ref{obs:faisability}
culminate into this final result on rational programs.
Suppose that a flow chart ${\cal C}$ 
representing the rational program ${\cal R}$ is partially known;
to be precise, suppose that the decision nodes
are known (that is, mathematical expressions
defining them are known) while the process nodes
are not. 
From Observation~\ref{obs:rationality},
each process node can be given by
a series of rational functions.
Determining each of those rational functions
can be achieved by solving 
an \textit{interpolation} or \textit{curve fitting} problem.

It follows that a performance prediction model
which can be depicted as a flow chart can also be depicted
as a piece-wise rational function and thus a rational program. 
For example, the MWP-CWP performance model
on top of \cuda\ (see Examples~\ref{ex:cuda} and \ref{ex:mwpcwp}) 
gives us an estimate of ${\cal E}$ from program, data, and hardware 
parameters and some low-level metrics. 
Thus, by determining rational functions describing these low-level metrics,
which are parameterized by program and data parameters,
we can fully determine a rational program estimating ${\cal E}$
which depends only on program and data parameters.

Recall that building such a rational
program ${\cal R}$ is our goal.  Once ${\cal R}$ is known, it can be
used at runtime (that is, when the program ${\cal P}$ is run
on specific input data) to compute optimal values for the program
parameters. This is exactly what is achieved in our tool implementing this technique. 
\fixed{Before exploring this tool we make one final remark. 
We assumed that the decision nodes in the flowchart of the rational program were known, 
however, we could relax this assumption.
Indeed, each decision node is given by
a series of rational functions.
Hence, those could be determined by
solving curve fitting problems as well.
We do not discuss this direction further
since this is not needed in our proposed
technique or implementation presented in the remainder of this paper.}{Could be removed? I agree that, while this a nice remark, this may only be clear to a mathematician.}

\subsection{KLARAPTOR: A Dynamic Optimization tool for CUDA}
\label{sec:mainresult_klaraptor}

\begin{figure*}[ht]
	\centering
	\includegraphics[width=\textwidth,clip,trim={1.5em, 1em, 5.5em, 1.2em}]{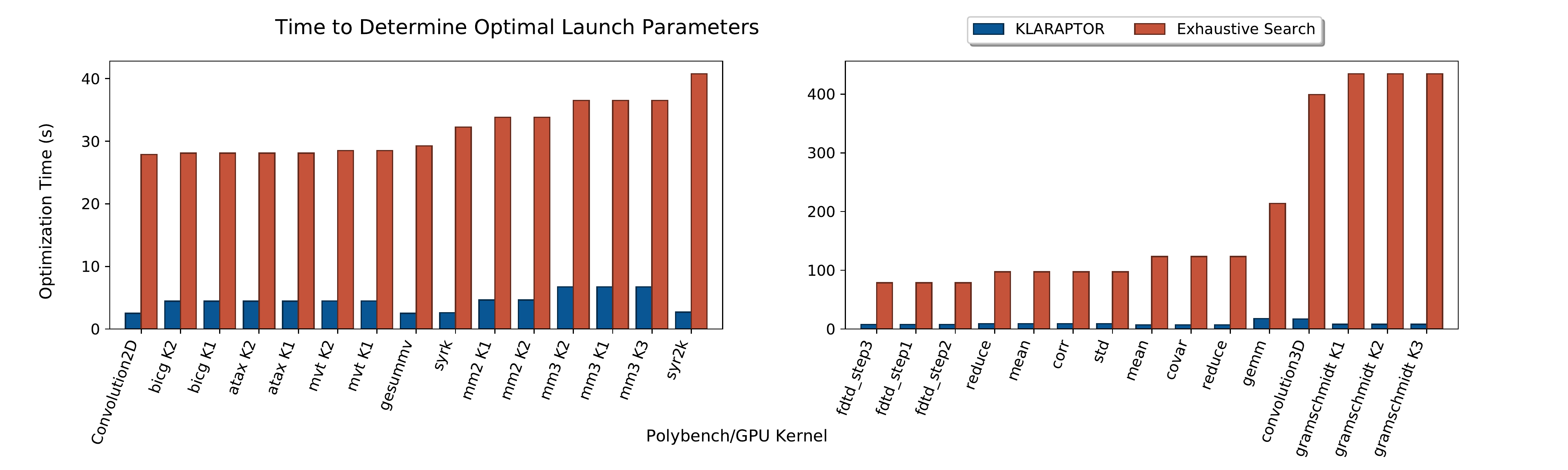}
	\caption{Comparing cumulative times to determine optimal launch parameters for data sizes $32 \leq N \leq 2048$ for each kernel in \texttt{Polybench/GPU}.}\label{fig:systemtimesall}
\end{figure*}

The theory of rational programs is put into practice for the {\cuda}
programming model by our tool KLARAPTOR.  KLARAPTOR is a compile-time
tool implemented using the LLVM Pass Framework and the {\mwpcwp}
performance model to dynamically choose a {\cuda} kernel's launch
parameters (thread block configuration) which optimize its
performance.  Most high-performance computing applications require
computations be as fast as possible and so kernel performance is
simply measured as its execution time.

As mentioned in Section~\ref{sec:intro}, 
thread block configurations drastically affect the running time of a kernel.
Determining optimal thread block configurations typically follows some heuristics, for example, 
constraining block size to be a multiple of 32 \cite{cuda2016best}. However, it is known
that the dimension sizes of a thread block, not only its total size, affect performance~\cite{DBLP:journals/tjs/TorresGL13,DBLP:conf/cascon/ChenCKMX15}.
Moreover, since thread block configurations are intimately tied to the size of data being operated on,
it is very unlikely that a static thread block configuration optimizes the performance 
of all data sizes. Our tool effectively uses a rational program to 
dynamically consider the particular data size of every kernel invocation and determines the thread block configuration
which minimizes the execution time of that particular invocation.
This is achieved in two main steps.
\begin{enumerate}
	\item At the compile-time of a {\cuda} program, its kernels are analyzed in order to
	build a rational program estimating the performance of each individual kernel under the {\mwpcwp} model. 
	Each rational program, written as code in the C language, is inserted into the code of the {\cuda} program
	so that it is called before the execution of the corresponding kernel.
	\item At runtime, immediately preceding the launch of a kernel, where data parameters have specific values, the rational program is evaluated to 
	determine the kernel launch parameters (thread block configuration) which optimize the performance of the kernel. The kernel
	is then launched using this thread block configuration. 
\end{enumerate} 

The practical benefits of KLARAPTOR are numerous, but in particular we are concerned with kernel performance
and programmer performance. Of course, by choice of optimal kernel launch parameters, 
the execution time of a kernel is minimized without any modifications to the kernel. 
By programmer performance we mean the efficiency of a programmer to produce 
optimal code. When a programmer is attempting to optimize a kernel, choosing optimal launch
parameters can either be completely ignored, 
performed heuristically, determined by trial and error, or determined by an exhaustive search.
The latter two options quickly become infeasible as data sizes grow large.
Regardless, any choice of optimal thread block configuration is likely to optimize
only a single data size. 

For KLARAPTOR to be practical, not only does the choice of optimal
kernel launch parameters need to be correct, but the two main steps
defined previously must also be performed in an way which is more
efficient than trial and error or exhaustive search.  Namely, the
compile-time analysis cannot overwhelm the compilation time and the
runtime decision of the kernel launch parameters cannot overwhelm the
kernel execution time.  For the former our analysis is performed very
quickly using only small data sizes whereas for the later the rational
program evaluation is fast and simple, maintaining a runtime history
to instantly provide results for future kernel launches.  The full
details of this process is presented in Section~\ref{sec:embodiment}.
In the remainder of this section we highlight the performance of
KLARAPTOR.

The \texttt{Polybench/GPU} benchmark suite provides an empirical
evaluation of our tool on a range of {\cuda} programs.  In
Fig.~\ref{fig:kernelperf2048} we have already seen that KLARAPTOR
accurately predicts the optimal or near-optimal thread block
configuration.  In Fig.~\ref{fig:systemtimesall} we compare the 
so-called system performance.  System performance describes the time
required to determine the optimal thread block configuration by either
(a) completely running KLARAPTOR, including rational program
construction, evaluation, and running the {\cuda} program over many
data sizes; or (b) invoking the kernel on all possible thread block
configurations in an exhaustive search over those same data sizes.
Evidently, using KLARAPTOR is orders of magnitude faster than an
exhaustive search while also maintaining the fact that it
dynamically adjusts for any data size.  More detailed results and
experimentation is presented in Section~\ref{sec:performance}.


\section{An Algorithm to Build and Deploy a Performance-Predicting Rational Program}
\label{sect:steps}

In this section, the notations and hypotheses are the same as in
Section~\ref{sec:foundations}.
Namely, ${\cal E}$ is a high-level performance metric for the 
multithreaded program ${\cal P}$ (e.g. execution time, memory consumption, or hardware occupancy),
${\bm L}$ is a set of low-level metrics given by the performance prediction model
or the particular platform of ${\cal P}$, and ${\bm P}$, ${\bm D}$, ${\bm H}$ are
sets of program, data, and hardware parameters, respectively. 

Let us assume that the values of ${\bm H}$ are known at compile-time of ${\cal P}$
while the values of ${\bm D}$ are known at runtime.
Further, let us assume that ${\bm P}$ and ${\bm D}$
take integer values. 
Hence the values of ${\bm P}$ belong to a finite set $F \subset \mathbb{Z}^p$.
%
%
That is to say, the possible values of
$\bm{P}$ are tuples of the form $(\pi_1, \ldots, \pi_p) \in F$,
with each $\pi_i$ being an integer.
Let us call such a tuple a \textit{configuration} of the program parameters.
Due to the nature of program parameters, those are not necessarily all
independent variables (i.e. a program parameter may depend on the value of
another program parameter). 
For example, in {\cuda}, the product of the dimension sizes
of a thread block is usually
\begin{inparaenum}[(i)]
\item a multiple of the warp size (32)
  and, \item bounded over by the maximum number of threads per block.
  \end{inparaenum}
\removed{even though the size of a thread block can vary between
1 and 1024, it should be a multiple of the warp size (32)
while an optimal choice is usually $2^b$ for 
$b\geq5$.} 

Following Section~\ref{sec:mainresult_rp}, we can compute a
rational program ${\cal R}$ estimating ${\cal E}$ as a function of
$\bm{D}$ and $\bm{P}$.  To do so, we shall first 
compute rational functions $g_1(\bm{D}, \bm{P})$,
\ldots, $g_{\ell}(\bm{D}, \bm{P})$, estimating
$L_1, \ldots, L_{\ell}$ respectively.
Once ${\cal R}$ has been computed, it can be used 
when ${\cal P}$ is being executed (thus at runtime)
in order to determine optimal values for $\bm{P}$,
on given values of $\bm{D}$.
The entire process is decomposed into six steps:
the first three occur at compile-time and the next three 
at runtime.

\begin{enumerate}
\item \textbf{Data collection}: 
\ifnewversion
In order to perform a curve fitting of the rational functions 
$g_i(\bm{D},\bm{P})$, $1 \leq i \leq \ell$, we require data points with which to perform
the fit. These are collected by 
\begin{inparaenum}[(i)]
\item selecting a set of points
$K$ which is a subset of the space of possible
values of $(\bm{D}$, $\bm{P})$; and
\item we execute the program ${\cal P}$, recording the values of
       the low-level performance metrics $\bm{L}$ as
       ${\bm V} = (V_1, \ldots, V_{\ell})$, at each point in $K$.
\end{inparaenum}
\else
Estimating $\cal{E}$ of the program for specific configurations of
program parameters and input data size. Through emulation, source-code
analysis, and use of existing performance models, this step gathers
the following statistics about the program and the target hardware:
\begin{enumerate}
\item architecture specific performance counters
\item runtime resources usage of the program
\item target hardware specific parameters
\end{enumerate}
\fi
\item \textbf{Rational function estimation}: 
\ifnewversion
For each low-level metric $L_i$ we use the set of points $K$ 
and the value $V_i$ measured for each point to estimate
the rational function $g_i(\bm{D},\bm{P})$. 
We observe that if the values of $V_i$ 
were known exactly (that is, without error) the rational function
$g_i(\bm{D}, \bm{P})$ could be determined exactly
via rational function interpolation.  In practice, however, these 
empirical values are likely to be noisy from profiling,
and/or numerical approximations.
Consequently, we determine a rational function
$\hat{g_i}(\bm{D}, \bm{P})$ which approximates
$g_i(\bm{D}, \bm{P})$.
\else
Using parameter estimation to fit a ``curve'' (rational function) to
the collected data for each of the independent variables of
interest. Examples of independent variables include number of compute
instructions, number of memory transactions, and number of
synchronization instructions.
\fi
\item \textbf{Code generation}: 
\ifnewversion
In order to generate the rational program ${\cal R}$, we 
proceed as follows:
\begin{enumerateshort}
\item[(i)] we convert the rational program representing 
${\cal E}$ into code, say in the C programming language, 
essentially encoding the CFG for computing $\cal{E}$;
\item[(ii)] we convert each
$\hat{g_i}(\bm{D}, \bm{P})$
into code, specifically sub-routines, estimating $L_i$; and
\item[(iii)] we include those sub-routines
into the code computing ${\cal E}$, which yields
the desired rational program ${\cal R}$.
\end{enumerateshort}
At this point the rational program ${\cal R}$ is fully determined.
\else
Devising a rational program to compute the high-level performance
metric of interest, $\cal{E}$. The generated rational program invokes
the rational functions generated in the earlier step.
\fi
\item \textbf{Rational program evaluation}: 
\ifnewversion
At the runtime of ${\cal P}$, the data parameters $\bm{D}$ are given
particular values, say ${\delta}_1, \ldots, {\delta}_d$.
For those specified values of $\bm{D}$ and for
all practically meaningful\footnote{``Practically
meaningful'' refers to the fact that the values of the
$\bm{P}$ are likely to be constrained by
the values $\bm{D}$.  For example,
if $P_1, P_2$ are the two dimension sizes of a two-dimensional
thread block of a {\cuda} kernel operating on a
square matrix of order $D_1$, then $P_1 P_2 \leq D_1^2$
is meaningful.} values of $\bm{P}$ from
the set $F$, we compute an estimate of ${\cal E}$ using ${\cal R}$.
The evaluation of ${\cal E}$ over so many different possible
program parameters is feasible for three reasons:
\begin{enumerateshort}
\item[(i)] The number of program parameters is small, typically $p \leq 3$,
      see Section~\ref{sec:embodiment};
\item[(ii)]  the set of meaningful values for ${\bm P}$ is small
	(consider that in {\cuda} the product of thread block dimensions should be a multiple of 32 less than 1024); and 
\item[(iii)]  the program ${\cal R}$ simply evaluates 
       a few formulae and thus runs almost instantaneously.
\end{enumerateshort}
\else
Running the rational program to evaluate $\cal{E}$ at different values
of program parameters, and the known hardware parameters.
\fi
\item \textbf{Selection of optimal values of program parameters}: 
\ifnewversion
When the search space of values of the program parameters
 $\bm{P}$ is large, a numerical optimization technique is required
 for this step. 
 However, as just explained, the number of evaluations to sort through
 is quite small and thus an exhaustive search is feasible. 
 However, due to inaccuracies in the performance prediction model
 being used, and in the estimation of the rational functions $g_i(\bm{D}, \bm{P})$
 as $\hat{g}_i(\bm{D}, \bm{P})$, there may be several configurations 
 which, up to some margin, optimize ${\cal E}$. 
Then, a secondary performance metric or some heuristic specific 
to the platform of ${\cal P}$ may be used to refine the choice 
of optimal configuration.
\else
Choosing the program parameters that minimize the estimated value of
$\cal{E}$ for the given program.
\fi
\item \textbf{Program execution}: 
\ifnewversion
Once an optimal configuration is selected, the
program ${\cal P}$ can finally be executed using this
configuration along with the values ${\delta}_1, \ldots, {\delta}_d$
of $\bm{D}$.

\else
Executing the program with the program parameters chosen in the
earlier step and the known hardware parameters for the target
architecture.
\fi
\end{enumerate}

\section{The Implementation of KLARAPTOR}
\label{sec:embodiment}

In this section we give an overview of the implementation of 
our previously presented technique (Section~\ref{sect:steps}) applied to {\cuda} 
using the {\mwpcwp} performance model. 
The KLARAPTOR tool is built in the C language, 
making use of the LLVM Pass Framework (see Section~\ref{sec:io-builder}) 
and the NVIDIA {\cupti} API (see Section~\ref{sec:data_collection}).
KLARAPTOR is freely available in source at 
\textcolor{navy}{\href{https://github.com/orcca-uwo/KLARAPTOR}{https://github.com/orcca-uwo/KLARAPTOR}}.

In the case of a {\cuda} kernel, the data parameters
specify the input data size.
In many examples this is a single parameter, say $N$,
describing the size of an array (or the order of a multi-dimensional array), 
the values of which are usually powers of $2$.
Program parameters describe
the kernel launch parameters, i.e. grid and thread block dimensions, 
and are also typically powers of $2$.
For example, a possible thread block configuration
may be $1024 \times 1 \times 1$ (a one-dimensional thread block),
or $16 \times 16 \times 2$ (a three-dimensional thread block).
Lastly, the hardware parameters are values specific to the 
target GPU, for example, memory bandwidth, the number of SMs available,
and their clock frequency.

\ifshortversion
This section is broken into subsections.
\fi
Sections~\ref{sec:annotation}, \ref{sec:preprocessing},
and \ref{sec:io-builder} are specific to our implementation and do not
\ifshortversion
correspond to any step of Section~\ref{sect:steps}.
\fi
\ifshortversion
The compile time steps 1 (data collection) and 2 (rational function estimation)
are reflected in Sections~\ref{sec:data_collection}
and \ref{sec:paramest}, while step 3 requires no additional explanation.
\fi
\ifshortversion
The runtime steps 4 (rational program evaluation),
5 (selection of optimal configuration),
and 6 (program execution)
are also straight forward from the algorithm in Section~\ref{sect:steps}.
\fi
Throughout this section we refer to the notion of a \textit{driver program}
as the code which includes the rational program and uses it to select
\ifshortversion
an optimal configuration.
\fi


%

\subsection{Annotating the source code}
\label{sec:annotation}
	Beginning with a {\cuda} program written in C/C++, we minimally annotate the host code
	to make it compatible with our \textit{pre-processor}.
	We take into account the following points:
	\begin{enumerateshort}
	\item[(i)] 
	the code targets CUDA Compute Capability 3.x or higher;
	\item[(ii)]
	there should be no {\cuda} runtime API calls as such calls will interfere
	with later {\cuda} driver API calls used by our tool, for example, \texttt{cudaSetDevice};
	\item[(iii)]
	the block dimensions and grid dimensions must be declared 
	as the typical {\cuda} {\dimThree} structs.
\end{enumerateshort}

For each kernel in the {\cuda} code, we add two 
pragmas, one defining the index of the kernel input arguments
corresponding to the data size $N$ and one specifying the dimension 
of the kernel (1, 2, or 3).
As an example, consider the code segment below for the \texttt{Sample} {\cuda} kernel
and the associated pragmas.
This kernel operates of a two-dimensional kernel of order $N$, 
making it a two-dimensional kernel.	
{
\begin{tcolorbox}
\scriptsize
\begin{verbatim}
#pragma kernel_info_size_param_idx_Sample = 1; 
#pragma kernel_info_dim_sample_kernel = 2;
__global__ void Sample (int *A, int N) { 
    int tid_x = threadIdx.x + blockIdx.x*blockDim.x;
    int tid_y = threadIdx.y + blockIdx.y*blockDim.y;
    ...
}
\end{verbatim}
\end{tcolorbox}
}

Lastly, for each kernel, the user must fill two formatted configuration 
files which follow \texttt{Python} syntax.
One specifies the constraints on the thread block configuration
while the other specifies the grid dimensions.
For example, for the 2D kernel \texttt{Sample} above, 
one could specify that its thread block configuration ($bx,by,bz$) must satisfy 
$bx < by^2$, $bx < N$ and $by < N$.
Similarly, the grid dimensions ($gx,gy,gz$), could be specified as
$gx=ceil\left[ \frac{N}{bx} \right]$ and 
$gy=ceil \left[ \frac{N}{by} \right]$. 

\subsection{Preprocessing}
\label{sec:preprocessing}
The preprocessor processes 
the annotated source replacing {\cuda} runtime API calls to driver 
API kernel launches. 
This step includes source code analysis in order to extract
the list of kernels, the list of kernel calls in the host code, 
and finally, the body of each kernel (which will be used for further analysis)
and building a so-called ``PTX lookup table'' to store
specific kernel information.
This table will be inserted into the so-called ``instrumented binary''---the
compiled {\cuda} program, augmented by our driver programs. 

	

\subsection{Input/Output Builder}
\label{sec:io-builder}
The Input/Output builder Pass, or IO-builder,
is a compiler pass written in the LLVM Pass Framework to
build the previously mentioned ``instrumented binary''.
This pass embeds IO mechanisms
to communicate with the driver program of each kernel at runtime.
There is one such function call embedded per kernel call.
Thus, at the runtime of the {\cuda} program being analyzed (step 6 of Section~\ref{sect:steps}),
an IO function is called before each kernel invocation to return six integers,
$(gx, gx, gz, bx, by, bz)$, the optimal kernel launch parameters.

The IO-builder pass goes through the following steps:
\begin{enumerate}[(i)]
\item obtain the LLVM IR (intermediate representation)
of the instrumented source code and find all {\cuda} driver API kernel calls;
\item relying on the annotated information for each kernel,
determine which variables in the IR contain the value of $N$ for a 
corresponding kernel call; and
\item insert a call to an IO function immediately before each kernel call
in order to pass the runtime values of $N$ to the
corresponding driver program and retrieve the optimal kernel
launch parameters.
\end{enumerate}
 
\subsection{Building a Driver Program: Data collection}
\label{sec:data_collection}
For the {\mwpcwp} model as well as our eventual rational
function estimation, 
we must collect data and statistics regarding certain performance counters and runtime
metrics (these are thoroughly defined in \cite{DBLP:conf/isca/HongK09}
and \cite{cuda2015}).
These metrics can be partitioned into three categories.

Firstly, \textit{architecture-specific performance counters} of a kernel,
basically, characteristics dictated by the CC
of the target device. Such characteristics can be obtained
at compile-time, as the target CC is specified at this time.
These characteristics include the number of registers used in each thread,
the amount of static shared memory 
per thread block, and the number of (arithmetic and memory)
instructions per thread.

Secondly, \textit{runtime-specific performance counters} that depend on
the behavior of the kernel at runtime.
This includes values impacted by memory access patterns, namely, 
the number of memory accesses per warp, the number of memory instructions in each thread,
and eventually, the total number of warps that are being executed.
We have developed a customized profiler using NVIDIA's EVENT API within the {\cupti} API
to collect the required runtime performance counters. 
The profiler is customized to collect exactly the information required for 
the {\mwpcwp} model and nothing else, allowing it
to be very lightweight and avoid the huge overheads of a typical profiler (e.g. NVIDIA's {\nvprof}~\cite{nvprofguide}).

Thirdly, \textit{device-specific parameters}, which
describe an actual GPU card, allow us to compute
a more precise performance estimate.
A subset of such parameters can be determined by microbenchmarking 
the device (see \cite{DBLP:journals/tpds/MeiC17} and \cite{DBLP:conf/ispass/WongPSM10}),
this includes the memory bandwidth, 
and the departure delay for memory accesses.
The remaining parameters can easily obtained
by consulting the vendor's guide \cite{cuda2019guide}, 
or by querying the device itself via the {{\cuda} driver API}.
Such parameters include the number of SMs on the card, 
the clock frequency of SM cores, and the instruction delay.

\subsection{Building a Driver Program: Rational function estimation}
\label{sec:paramest}
Using the runtime data collected in the previous step, 
we look to determine the rational functions
$\hat{g}_i(\bm{D},\bm{P})$ (see Section~\ref{sect:steps})
which estimate the low-level metrics or other intermediate 
values in the performance model. For simplicity of
discussion, we replace the parameters $\bm{D}$ and $\bm{P}$
with the generic variables $X_1,\ldots,X_n$.

A rational function is simply a fraction of two polynomials. With
a \textit{degree bound} (an upper limit on the exponent) on each
variable $X_k$ in the numerator and the denominator,
say $u_k$ and $v_k$, respectively, these polynomials can be defined up to
some \textit{parameters} (using the language of parameter estimation), namely the coefficients of the polynomials,
say $\alpha_1,\dots,\alpha_i$ and $\beta_1,\dots,\beta_j$.
Through analysis of the MWP-CWP model these degree bounds,
and thus $i$ and $j$, are relatively small.
 
\scalebox{0.88}{
\begin{minipage}{\linewidth}
\begin{align*}
f_b(X_1,\dots,X_n) &= p_b(X_1,\dots,X_n) \; / \; q_b(X_1,\dots,X_n) \\
&= \frac{\alpha_1\cdot(X_1^0\cdots X_n^0) \;+\; 
	\dots \;+\;  \alpha_i\cdot(X_1^{u_1}\cdots X_n^{u_n})}{\beta_1\cdot(X_1^0\cdots X_n^0) \;+\;
	\dots \;+\; \beta_j\cdot(X_1^{v_1}\cdots X_n^{v_n})}
\end{align*}
\end{minipage}
}

We perform a parameter estimation (for each rational function)
on the coefficients $\alpha_1, \ldots, \alpha_i, \beta_1, \ldots, \beta_j$. 
to determine the rational function. 
This is a relatively simple linear regression 
which can be solved as an over-determined
system of linear equations,
say by the method of linear least squares.
%

However, our system is constructed from the evaluation 
of monomial terms, resulting in essentially a Vandermonde matrix.
Such matrices are very ill-conditioned.
Since we are interested in using our fitted model for 
extrapolation (i.e. estimating program parameters for new data parameters)
any numerical errors in the model fitting will 
grow very quickly \cite{corless2013graduate}. 
Thus, our solution must be as numerically stable
as possible. Moreover, since we have constructed the system 
from monomials, it suffers from \textit{multicollinearity} (see \cite[Chapter 23]{gentle2012handbook})
and can become rank-deficient.
The typical method to solve the system using QR-factorization 
is thus impossible and we use the computationally more intensive yet
more numerically stable method of \textit{singular value decomposition}
(for details see \cite[Chapter 4]{corless2013graduate}).

Our implementation uses the LAPACK (Linear Algebra PACKage) library \cite{userguide:lapack} for linear algebra
and the Basic Polynomial Algebra Subprograms (BPAS) library \cite{bpasweb}
for efficient rational function and polynomial implementations.

\section{Experimentation}
\label{sec:performance}

In this section we explore the performance of KLARAPTOR 
by applying it to the {\cuda} programs of the 
\texttt{Polybench/GPU} benchmark suite \cite{grauer2012auto}.
Throughout this section data was collected using a GTX 1080Ti. 

We note here that many of the kernels in this
suite perform relatively low amounts of work;
they are best suited to being executed many times
from a loop in the host code.
These kernels with extremely fast execution
create a large variance in relative timings
from trial to trial where the absolute time
scale is on the order of a fraction of a millisecond.
Performance is further varied by dynamic
frequency and voltage scaling 
despite setting the performance mode to peak performance.

\begin{figure*}
\centering
\includegraphics[width=\textwidth,trim={2em, 0em, 1.5em, 0em}, clip]{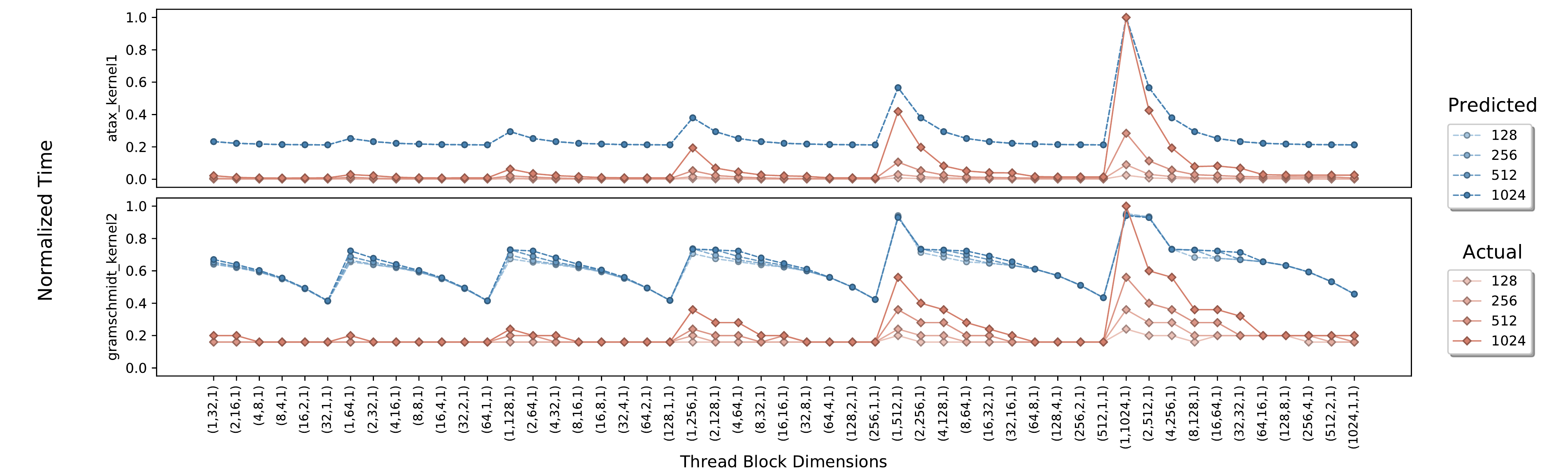}
\caption{The predicted and actual execution time of three kernels for various input sizes. The trends show that predicted minima occur at actual minima.}\label{fig:redbluecurves}
\end{figure*}

However, when it comes to choosing thread block configurations
which give optimal execution times, we need only to look at trends
and relative performance. Fig.~\ref{fig:redbluecurves} shows
these trends for three different kernels: \texttt{atax}, \texttt{corr}, and \texttt{gramschmidt}.
While the predicted values are not exact, they do show that the overall shape is 
accurate and that predicted minima align with actual minima in execution time.

\begin{table}[htb]
\centering
\scriptsize
\caption{Comparing KLARAPTOR against exhaustive search for thread block configuration choice of kernels in \texttt{Polybench/GPU}.}\label{table:polybenchConfigCompare}
\begin{tabular}{|l|r||r|r||r|r|}
\hline
\multirow{2}{*}{Kernel} & \multirow{2}{*}{N} & Chosen & {Time} & Best$\ \ $ & {Time} \\
	& & Config. & (ms) & Config. &  (ms) \\
\hline
\multirow{2}{*}{{\scriptsize Convolution2D}} & 8192 & 32, 16 &   2.42 & 32, 4 &   2.34 \\
 & 16384 & 32, 16 &  11.15 & 128, 1 &   9.34 \\
\hline
\multirow{2}{*}{{\scriptsize mm2 K1}} & 256 & 32, 16 &   0.20 & 512, 1 &   0.14 \\
 & 512 & 32, 16 &   0.97 & 32, 2 &   0.80 \\
\hline
\multirow{2}{*}{{\scriptsize mm2 K2}} & 256 & 32, 16 &   0.20 & 16, 2 &   0.17 \\
 & 512 & 32, 16 &   0.96 & 32, 2 &   0.80 \\
\hline
\multirow{2}{*}{{\scriptsize convolution3D}} & 512 & 32, 16 &   0.03 & 32, 8 &   0.03 \\
 & 1024 & 32, 16 &   0.07 & 32, 8 &   0.07 \\
\hline
\multirow{2}{*}{{\scriptsize mm3 K1}} & 1024 & 32, 16 &   7.48 & 32, 8 &   7.44 \\
 & 2048 & 32, 16 &  60.71 & 32, 8 &  61.03 \\
\hline
\multirow{2}{*}{{\scriptsize mm3 K2}} & 1024 & 32, 16 &   7.46 & 32, 8 &   7.57 \\
 & 2048 & 32, 16 &  60.66 & 128, 8 &  56.58 \\
\hline
\multirow{2}{*}{{\scriptsize mm3 K3}} & 1024 & 32, 16 &   7.44 & 32, 8 &   7.43 \\
 & 2048 & 32, 16 &  59.70 & 32, 8 &  55.80 \\
\hline
\multirow{2}{*}{{\scriptsize atax K1}} & 1024 & 32, 16 &   0.89 & 8, 4 &   0.20 \\
 & 2048 & 64, 8 &   1.74 & 16, 2 &   0.36 \\
\hline
\multirow{2}{*}{{\scriptsize atax K2}} & 1024 & 32, 16 &   0.28 & 128, 4 &   0.25 \\
 & 2048 & 64, 8 &   0.56 & 32, 1 &   0.49 \\
\hline
\multirow{2}{*}{{\scriptsize bicg K1}} & 4096 & 128, 8 &   1.74 & 256, 1 &   1.04 \\
 & 8192 & 256, 4 &   2.81 & 128, 1 &   2.21 \\
\hline
\multirow{2}{*}{{\scriptsize bicg K2}} & 4096 & 128, 8 &   8.43 & 32, 1 &   0.85 \\
 & 8192 & 256, 4 &  13.19 & 16, 2 &   4.30 \\
\hline
\multirow{2}{*}{{\scriptsize corr}} & 1024 & 32, 16 & 778.66 & 4, 1 & 621.80 \\
 & 2048 & 64, 8 & 2966.50 & 4, 1 & 3632.81 \\
\hline
\multirow{2}{*}{{\scriptsize mean}} & 1024 & 32, 16 &   0.41 & 4, 1 &   0.28 \\
 & 2048 & 64, 8 &   0.58 & 32, 1 &   0.55 \\
\hline
\multirow{2}{*}{{\scriptsize reduce}} & 1024 & 32, 16 &   0.04 & 1, 8 &   0.22 \\
 & 2048 & 32, 16 &   0.13 & 1, 8 &   0.83 \\
\hline
\multirow{2}{*}{{\scriptsize std}} & 1024 & 32, 32 &   0.88 & 4, 1 &   0.28 \\
 & 2048 & 64, 8 &   0.59 & 32, 1 &   0.56 \\
\hline
\multirow{2}{*}{{\scriptsize covar}} & 1024 & 32, 16 & 789.44 & 4, 1 & 622.18 \\
 & 2048 & 64, 8 & 2971.68 & 4, 1 & 3656.70 \\
\hline
\multirow{2}{*}{{\scriptsize mean}} & 1024 & 32, 16 &   0.40 & 4, 1 &   0.29 \\
 & 2048 & 64, 8 &   0.58 & 128, 1 &   0.55 \\
\hline
\multirow{2}{*}{{\scriptsize reduce}} & 1024 & 128, 8 &   0.01 & 32, 1 &   0.01 \\
 & 2048 & 256, 4 &   0.01 & 64, 1 &   0.01 \\
\hline
\multirow{2}{*}{{\scriptsize fdtd\_step1}} & 512 & 32, 16 &   0.02 & 32, 8 &   0.01 \\
 & 1024 & 32, 16 &   0.05 & 64, 8 &   0.04 \\
\hline
\multirow{2}{*}{{\scriptsize fdtd\_step2}} & 512 & 32, 16 &   0.02 & 64, 8 &   0.01 \\
 & 1024 & 32, 16 &   0.05 & 64, 8 &   0.04 \\
\hline
\multirow{2}{*}{{\scriptsize fdtd\_step3}} & 512 & 32, 16 &   0.02 & 32, 8 &   0.01 \\
 & 1024 & 32, 16 &   0.06 & 32, 8 &   0.05 \\
\hline
\multirow{2}{*}{{\scriptsize gemm}} & 1024 & 32, 16 &   7.49 & 32, 8 &   7.34 \\
 & 2048 & 32, 16 &  59.68 & 32, 8 &  61.43 \\
\hline
\multirow{2}{*}{{\scriptsize gesummv}} & 1024 & 32, 16 &   2.39 & 8, 4 &   0.34 \\
 & 2048 & 64, 8 &   4.66 & 16, 2 &   0.73 \\
\hline
\multirow{2}{*}{{\scriptsize gramschmidt K1}} & 512 & 512, 1 &   0.02 & 8, 1 &   0.02 \\
 & 1024 & 512, 1 &   0.04 & 1, 1 &   0.03 \\
\hline
\multirow{2}{*}{{\scriptsize gramschmidt K2}} & 512 & 1024, 1 &   0.01 & 32, 1 &   0.01 \\
 & 1024 & 32, 16 &   0.01 & 16, 1 &   0.01 \\
\hline
\multirow{2}{*}{{\scriptsize gramschmidt K3}} & 512 & 256, 4 &   0.26 & 16, 1 &   0.20 \\
 & 1024 & 32, 16 &   0.66 & 32, 1 &   0.54 \\
\hline
\multirow{2}{*}{{\scriptsize mvt K1}} & 4096 & 128, 8 &   6.64 & 32, 1 &   0.85 \\
 & 8192 & 256, 4 &  12.70 & 16, 2 &   4.32 \\
\hline
\multirow{2}{*}{{\scriptsize mvt K2}} & 4096 & 128, 8 &   1.39 & 32, 1 &   1.03 \\
 & 8192 & 256, 4 &   2.71 & 64, 1 &   2.19 \\
\hline
\multirow{2}{*}{{\scriptsize syr2k}} & 1024 & 32, 16 &  59.59 & 8, 8 &  58.64 \\
 & 2048 & 32, 16 & 468.96 & 4, 8 & 670.23 \\
\hline
\multirow{2}{*}{{\scriptsize syrk}} & 1024 & 32, 16 &  30.42 & 16, 8 &  30.22 \\
 & 2048 & 32, 16 & 240.84 & 16, 8 & 247.60 \\
\hline
\end{tabular}
\end{table}

Table~\ref{table:polybenchConfigCompare} 
provides experimentation data for every kernel in the 
\texttt{Polybench/GPU} benchmark suite. Namely, this table compares 
the execution times of the thread block configuration chosen by KLARAPTOR against
the optimal thread block configuration found though exhaustive search. 
The table shows a couple of data sizes in order to highlight that the best configuration 
can change for different input sizes.




\section{Conclusions and Future Work}
\label{sec:conclusion}

The performance of a single {\cuda} program can vary wildly
depending on the target GPU device, the input data size, 
and the kernel launch parameters. Moreover, a thread block
configuration yielding optimal performance for a particular data size or a particular
target device will not necessarily be optimal for
a different data size or different target device. 
In this paper we have presented the KLARAPTOR tool for 
determining optimal {\cuda} thread block configurations 
which is adaptive to each kernel invocation and input data, 
allowing for dynamic data-dependent performance and portable performance.
This tool is based upon our technique of encoding a 
performance prediction model as a rational program.
The process of constructing such a rational program is a fast 
and automatic compile-time
process which occurs simultaneously to compiling
the {\cuda} program by use of the LLVM Pass framework. 
Our tool was tested using the kernels of the \texttt{Polybench/GPU} benchmark suite
with good results.

That same experimentation has lead us to consider 
the limitations our chosen performance prediction model, {\mwpcwp}.
Recently, the author of \cite{DBLP:conf/ppopp/Volkov18} and \cite{Volkov:EECS-2016-143}
has suggested a new GPU performance 
model which relies on Little's law;
it measures concurrency as a product of latency and throughput.
This model considers both warp and instruction concurrency while 
the other models such as \cite{cuda2019guide}, \cite{DBLP:conf/isca/HongK09}, \cite{DBLP:conf/ppopp/SimDKV12}, and \cite{Baghsorkhi:2010:APM},
consider only warp concurrency. 
The author's analysis of those models suggests their limitation 
is the significant underestimation of occupancy when arithmetic intensity 
(the number of arithmetic instructions per memory access) is intermediate. 
This is exactly the type of kernels on which KLARAPTOR underperforms.
In future work we look to apply an improved performance prediction
in order to achieve even better results.


%
%

\bibliographystyle{IEEEtran}
\bibliography{IEEEabrv,CRD,multi+arbitrary-precision,BPAS,patents,gpu_performance_models}

\end{document}